\documentclass[a4paper]{jpconf}
\usepackage{graphicx}

\newcommand{\be}{\begin{eqnarray}}
\newcommand{\ee}{\end{eqnarray}}
\def\nue{{\nu_e}}
\def\anue{{\bar\nu_e}}
\def\numu{{\nu_{\mu}}}
\def\anumu{{\bar\nu_{\mu}}}
\def\nutau{{\nu_{\tau}}}

\newcommand{\ma}{\Delta m^2_{31}}

\newcommand{\sch}{\sin^2 \theta_{13}}

\newcommand{\sa}{\sin^2 \theta_{23}}
\newcommand{\sta}{\sin^22 \theta_{23}}

\newcommand{\scht}{\sin^2 \theta_{13}{\mbox {(true)}}}

\newcommand{\sat}{\sin^2 \theta_{23}{\mbox {(true)}}}

\newcommand{\sig}{$3\sigma$}

\begin{document}
\title{What we can learn from atmospheric neutrinos}

\author{Sandhya Choubey}

\address{Harish-Chandra Research Institute, Chhatnag Road, Jhunsi, 
Allahabad 211019, India}

\ead{sandhya@mri.ernet.in}

\begin{abstract}
Physics potential of future measurements
of atmospheric neutrinos is explored. 
Observation of 
$\Delta m^2_{21}$ driven sub-dominant 
effects and $\theta_{13}$ driven 
large matter effects in atmospheric neutrinos can be 
used to study the 
deviation of $\theta_{23}$ from maximality and its octant.
Neutrino mass hierarchy can be determined extremely well due 
to the large matter effects. New physics can be constrained
both in standard atmospheric neutrino experiments as well as 
in future neutrino telescopes.
\end{abstract}

%%%%%%%%%%%%%%%%%%%%%%%%%%%
\section{Introduction}
%%%%%%%%%%%%%%%%%%%%%

Atmospheric neutrinos observed in the Super-Kamiokande (SK) 
experiment provided the 
first unambiguous signal for neutrino flavor oscillations 
\cite{Ashie:2005ik}. The observed zenith angle and energy 
dependent depletion of atmospheric $\numu/\anumu$ in SK
can be explained only 
%together with no depletion of $\nue/\anue$ by 
by $\numu$-$\nutau$ oscillations with $\ma=2.1 \times 10^{-3}$ eV$^2$ 
and almost maximal mixing angle, $\sta=1$. The results of the 
SK experiment was subsequently corroborated by the 
MACRO and Soudan-2 atmospheric neutrino experiments %\cite{atmother}
and more recently by the K2K and MINOS long baseline (LBL)
experiments. %\cite{lbl}. 
In this talk we will expound the 
physics potential of 
future atmospheric neutrino experiments using larger and 
better detectors. 
%Here we will discuss what more we can 
%learn by doing atmospheric neutrino experiments using larger and 
%better detectors in the future. 
%While we discuss only the potential 
%of large water and/or magnetized detectors, similar results can be 
%obtained using large (magnetized) liquid argon detectors.

%%%%%%%%%%%%%%%%%%%%%%%%%%%%%%%%%%%%%%%%%%%%%%%%%%%%%%%%%%
\section{Confirming oscillations of atmospheric neutrinos}
%%%%%%%%%%%%%%%%%%%%%%%%%%%%%%%%%%%%%%%

The ``smoking gun'' signal for $\numu$-$\nutau$ oscillations 
is the observation of the characteristic ``dip'' in $L/E$,  
predicted by neutrino flavor mixing. Although the analysis 
of the $L/E$ binned atmospheric neutrino 
data in SK has been found to support the 
oscillation hypothesis \cite{lbye}, it would be worthwhile to 
make an unambiguous check using a detector with better 
$E$ and $L$ resolution. 
This can be done in
large magnetized detectors, such as the proposed ICAL 
detector at the India-based Neutrino Observatory (INO) 
\cite{Athar:2006yb}. 
%which has good resolution for both $E$ and $L$. 
Analysis of results obtained from detailed simulations by 
the INO collaboration show that oscillations can be confirmed with 
a significant C.L. with just 250 kTy data.

%%%%%%%%%%%%%%%%%%%%%%%%%%%%%%%%%%%%%%%%%%%%%%%%%%%%%%%%%%
\section{Precision measurement of $\ma$ and $\sa$}
%%%%%%%%%%%%%%%%%%%%%%%
Both $\ma$ and $\sta$ are expected to be measured very 
accurately by the forthcoming LBL experiments.
A statistical analysis of the combined data set with
five years of 
running of MINOS, ICARUS, OPERA, T2K and NO$\nu$A 
each, reveals that $\ma$ and $\sa$ could be measured with 
a spread of 4.5\% 
and 20\% respectively at \sig{} \cite{huber10}.
The future prospective data from water Cerenkov atmospheric 
neutrino experiments with a statistics 20 times the current 
SK statistics could measure $\ma$ and $\sa$ with a spread of 
$\sim 17\%$ and $\sim 24\%$ respectively \cite{Gonzalez-Garcia:2004cu}.
A large magnetized iron calorimetric detector
such as the proposed
INO detector ICAL \cite{Athar:2006yb}, 
could use atmospheric neutrinos to
measure $\ma$ 
and $\sa$ within 10\% and 30\% respectively 
at \sig{} with a statistics of
250 kTy \cite{Athar:2006yb}. 
%accuracy comparable to that expected from the combined 
%data from the LBL experiments
%of 3\% and 18\% respectively at \sig{} 
%\cite{inootherparams}.

%%%%%%%%%%%%%%%%%%%%%%%%%%%%%%%%%%%%%%%%%%%%%%%%
\section{Atmospheric neutrino experiments: Subdominant effects}
\label{sec:atm_future}
%%%%%%%%%%%%%%%%%%%%%%%

The effect of the sub-dominant terms 
in the Super-Kamiokande (SK) 
atmospheric neutrino data is not yet at the 
statistically significant level.
However, the sub-dominant terms, if observed in 
a future high statistics atmospheric neutrino experiment,
can be used to give information on: 
\begin{itemize}
\item Deviation of $\theta_{23}$ from its maximal value
\item Octant of $\theta_{23}$
\item $sgn(\Delta m_{31}^2)$
\end{itemize} 

\noindent
Assuming a constant density for the earth matter, 
the excess of electron type events in a water Cerenkov 
experiment such as SK is given by \cite{Peres:1999yi,Peres:2003wd}
%\footnote{The interference effects coming from the 
%core-mantle interface inside the earth can be important and 
%have been discussed in 
%\cite{Petcov:1998su,Akhmedov:1998ui,Bernabeu:2001xn,Bernabeu:2003yp}.}
\begin{eqnarray}
\frac{N_e}{N_e^0} - 1 &\simeq& 
\sin^22\theta_{12}^M
\sin^2\left(\frac{(\Delta m_{21}^2)^M L}{4E}\right ) 
\times (r\cos^2\theta_{23} - 1)\\
&+& \sin^22\theta_{13}^M
\sin^2\left(\frac{(\ma)^M L}{4E}\right )
\times (r\sin^2\theta_{23} - 1)\\
&+& \sin\theta_{23}\cos\theta_{23} ~ r ~ Re
\bigg[A_{13}^* A_{12} \exp(-i\delta_{CP})\bigg]~,
\label{eq:electronexcess}
\end{eqnarray}
where $L$ is the baseline, $E$ is the energy of the neutrino,
$r=N_e/N_\mu$, $N_e$ and $N_\mu$ being the number of 
e and $\mu$ events respectively
in the detector in absence of 
oscillations and $\theta_{12}^M$, $\theta_{13}^M$, 
$(\Delta m_{21}^2)^M$ and $(\ma)^M$ are the mixing angle and mass squared 
differences in matter. 
%The quantities $A_{13}$ and $A_{12}$ are

\begin{enumerate}
\item The first term in Eq. (\ref{eq:electronexcess})
is the $\Delta m_{21}^2$ driven oscillation term --
which is obviously more important for the sub-GeV neutrino sample. 
Since $r\simeq 0.5$ in the sub-GeV regime, this term brings an 
excess (depletion) of sub-GeV electron events if $\theta_{23} < \pi/4$
($\theta_{23} > \pi/4$). It can thus be used to study the maximality 
and octant of $\theta_{23}$
through the sub-GeV electron sample 
\cite{Peres:1999yi,Gonzalez-Garcia:2004cu}.

\item The second term is the $\theta_{13}$ driven oscillation term. 
Being dependent on $\sin^2\theta_{23}$,
this term goes in the opposite direction 
to the first term. Therefore for sub-GeV neutrinos, larger 
$\theta_{13}$ would imply that the effect of the first term 
would get negated by this term. However for multi-GeV neutrinos,
there will be large matter effects inside the earth and 
this term dictates the electron excess.
The $\sin^2\theta_{23}$ 
dependence of this term could then be used to study the maximality 
and octant of $\theta_{23}$ through the 
multi-GeV electron sample (see also \cite{Huber:2005ep}).
Since matter effects bring in sensitivity to the $sgn(\Delta m_{31}^2)$, 
this term can be used to study the mass hierarchy.

\item The last term is the ``interference'' term 
\cite{Peres:2003wd}, which depends on 
$\delta_{CP}$. The effect of this term could be to dilute the 
effect of the first two terms and spoil the sensitivity of the 
experiment. However, being directly dependent on $\delta_{CP}$, 
this term also brings in some sensitivity to the CP phase itself
\cite{Peres:2003wd,Fogli:2005cq}.
\end{enumerate}
%\noindent
The depletion of the muon events in the limit of $\Delta m_{21}^2=0$ 
is given by\footnote{The approximation of taking a vanishing 
$\Delta m_{21}^2$ has been made in Eq. (\ref{eq:muondeficit})
only for the sake of simplicity, since the main subdominant effect
in the muon neutrino channel comes from earth matter effects, 
which are large for multi-GeV neutrinos for which $\Delta m_{21}^2$
dependence is less importance. The results presented in the 
later sections have been obtained
using the full numerical solution of the three-generation 
equation of the atmospheric neutrinos.}
\be
1 - \frac{N_\mu}{N_\mu^0} &=& 
(P^1_{\mu\mu} + P^2_{\mu\mu}) + (P^3_{\mu\mu})^\prime \sin^2\theta_{23}
(\sin^2\theta_{23} - \frac{1}{r})~,
\ee
%where,
\be
P^1_{\mu\mu} &=& \sin^2 \theta^M_{13} {\sin^2 2\theta_{23}} \sin^2
{\left[({A}+\Delta m_{31}^2) - {(\Delta m_{31}^2)}^M\right]L 
\over 8E}~, \\
P^2_{\mu\mu} &=& \cos^2 \theta^M_{13} {\sin^2 2\theta_{23}} \sin^2
{\left [({A}+\Delta m_{31}^2 ) + {(\Delta m_{31}^2)}^M\right ]L \over 8E} ~,
\\
(P^3_{\mu\mu})^\prime 
&=& \sin^2 2\theta^M_{13} \sin^2
{{(\Delta m_{31}^2)}^M L\over 4E} ~,
\label{eq:muondeficit}
\ee
where $A=2\sqrt{2}G_FN_eE$ is the matter potential.
For very 
small values of $\theta_{13}$, there is very little matter effect and 
we can see that $P^{2}_{\mu\mu}$ 
is the dominant term in the survival probability.
Since this term depends on $\sin^2 2\theta_{23}$ we do not 
expect octant sensitivity in absence of matter effects from 
experiments probing the $P_{\mu\mu}$ channel alone. However, 
if $\theta_{13}$ is not too small, neutrinos which travel through 
large baselines and hence large matter densities inside
the earth, undergo large matter effects. The mixing angle 
$\theta_{13}^M$ increases in matter and the third term 
$(P_{\mu\mu}^3)^\prime$ 
becomes important as well. Since this term has a strong dependence on  
$\sin^2\theta_{23}$ rather than $\sin^2 2\theta_{23}$, we expect 
the $P_{\mu\mu}$ channel to develop sensitivity to the octant of 
$\theta_{23}$ in presence of large matter effects 
\cite{Choubey:2005zy}.
Probing matter effects in the resultant muon signal
in the detector will also provide us with information on  
the neutrino mass hierarchy 
\cite{hierarchy,Petcov:2005rv}.
%Bernabeu:2001xn,Palomares-Ruiz:2004tk,
%Indumathi:2004kd,Gandhi:2004md,Gandhi:2004bj,Gandhi:2005wa,Petcov:2005rv}

Since matter 
effects are large for higher energy neutrinos, we expect that 
multi-GeV atmospheric $\numu/\anumu$ events can be used for 
this purpose. 
However, unlike in the case for matter effects in the $P_{\mu e}$ 
channel, both the magnitude and sign of the earth 
matter effects in the $P_{\mu\mu}$ channel depends 
crucially on $L$ and $E$. 
The largest effect of earth matter comes for neutrino travelling 
$L\simeq 7000$ km with $E\sim 5$ GeV. 
The matter effects changes sign rapidly with $L$ and $E$ -- 
with $\Delta(P_{\mu\mu})<0$ and $\Delta(P_{\mu\mu})>0$
at the maximum and minimum respectively of $P_{\mu\mu}$. 
%This is illustrated in Fig. 
%\ref{Fig:SnuM:p22L} \cite{Choubey:2005zy}, 
%which shows the difference between 
%the ratio of 
%``upward going'' muon events to the ``downward going'' muon events
%for the atmospheric 
%neutrinos ($U_N/D_N$) and  antineutrinos ($U_A/D_A$),
%for a large magnetized iron detector, such as the 
%proposed ICAL detector at the India-based Neutrino 
%Observatory (INO) \cite{Athar:2006yb}. 
%The normal mass hierarchy is assumed and the results are shown 
%for different energy and zenith angle bins. 
%Since for a given mass hierarchy, large matter effects appear 
%in only either the neutrino or the antineutrino channel, this 
%difference in the ratio between the upward and downward events
%gives the net earth matter effects encountered by the 
%atmospheric neutrinos. We can note from the figure that
%the largest effect of earth matter comes for neutrino travelling 
%$L\simeq 7000$ km with $E\sim 5$ GeV. We can also see that 
%the net matter effects changes sign rapidly with $L$ and $E$. 
Thus, in order to see the matter effects one needs to bin the 
data judiciously both in energy and zenith angle. 
%Finally, we can see 
%that $\Delta P_{\mu\mu}$ depends on the value of $\theta_{23}$.

Very good energy and zenith angle detector resolution is 
expected for the magnetized iron calorimeters. 
Therefore, fine binning 
would allow such detectors to observe matter effects in the 
muon signal. Since large matter effects appear only in either 
the neutrino or the antineutrino channel, the magnetic
field which allows for charge discrimination, further 
helps these type of detectors to observe earth matter effects 
in the muon channel. However, unless the iron plates of the 
detector are thin enough, it would not be possible to detect
electrons in these type of detectors. The current INO-ICAL 
design does not allow for it and therefore would observe 
muon events only. Another restriction for these detectors 
come from the relatively higher threshold, which allows for the 
detection of only multi-GeV $\numu/\anumu$.

Water Cerenkov detectors have the advantage that sub-GeV
neutrinos can be detected. However, the energy resolution 
is worse than that for iron calorimeters. For the results 
shown here, the data is binned in sub-GeV and multi-GeV bins
and therefore the matter effects in the 
$P_{\mu\mu}$ channel get largely averaged out. This means that 
one would see very small residual matter effects in the multi-GeV 
muon sample. However, matter effects in the $P_{\mu e}$ channel 
do not change sign over most of the relevant range of $E$ and 
$L$ in the multi-GeV regime. Therefore, multi-GeV electron sample
has large matter effects and can be used to study the 
deviation of $\theta_{23}$ from maximality and its octant as well 
as the neutrino mass hierarchy.

%%%%%%%%%%%%%%%%%%%%%%%%%%%%%%%%%%%%%%%%%%%%%%%%%%%%%%%%%%
\section{Is the mixing angle $\theta_{23}$ maximal?}
%%%%%%%%%%%%%%%%%%%%%%%%%%%%%%%%%%%%

%%%%%%%%%%%%%%%%%%%%%%%%%%%%%%%%%%%%%%%%%%%%%%%%%%%%%%%%%%
\begin{figure}
\includegraphics[width= 5.0cm, height=5.2cm]{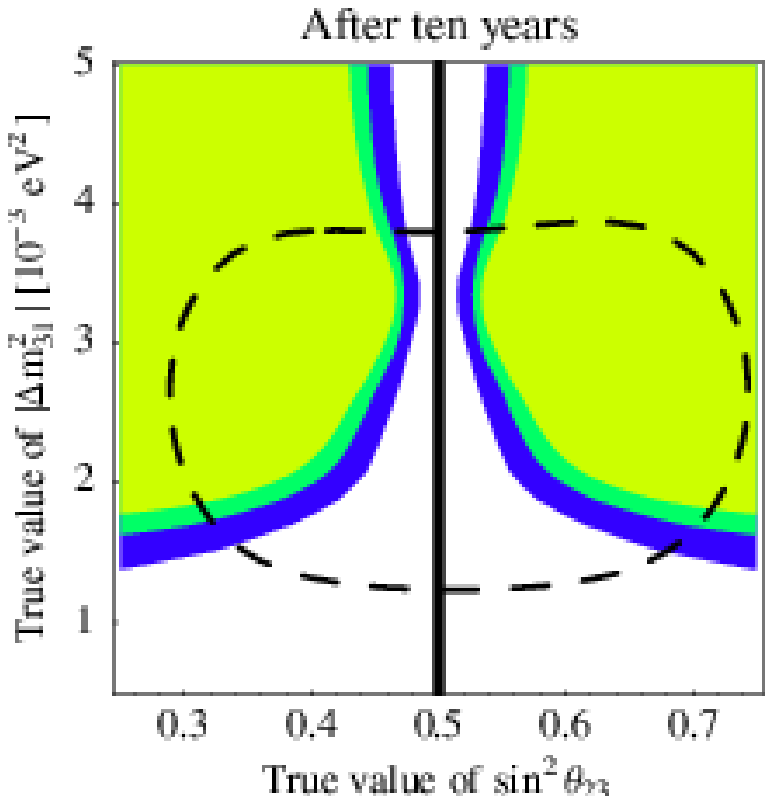}%\\
\vglue -4.8cm \hglue 5.0cm
\includegraphics[width= 5.0cm, height=5.0cm]{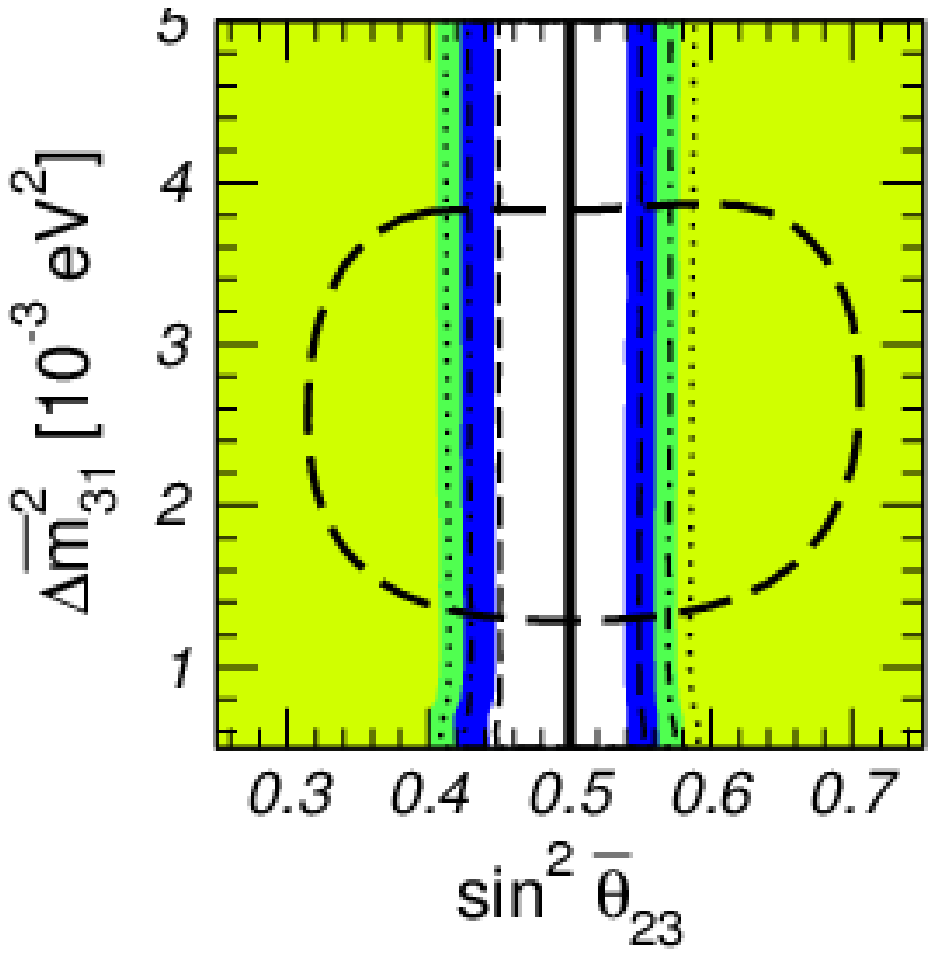}\\
\vglue -5.9cm \hglue 10.5cm
\includegraphics[width= 5.0cm, height=5.2cm]{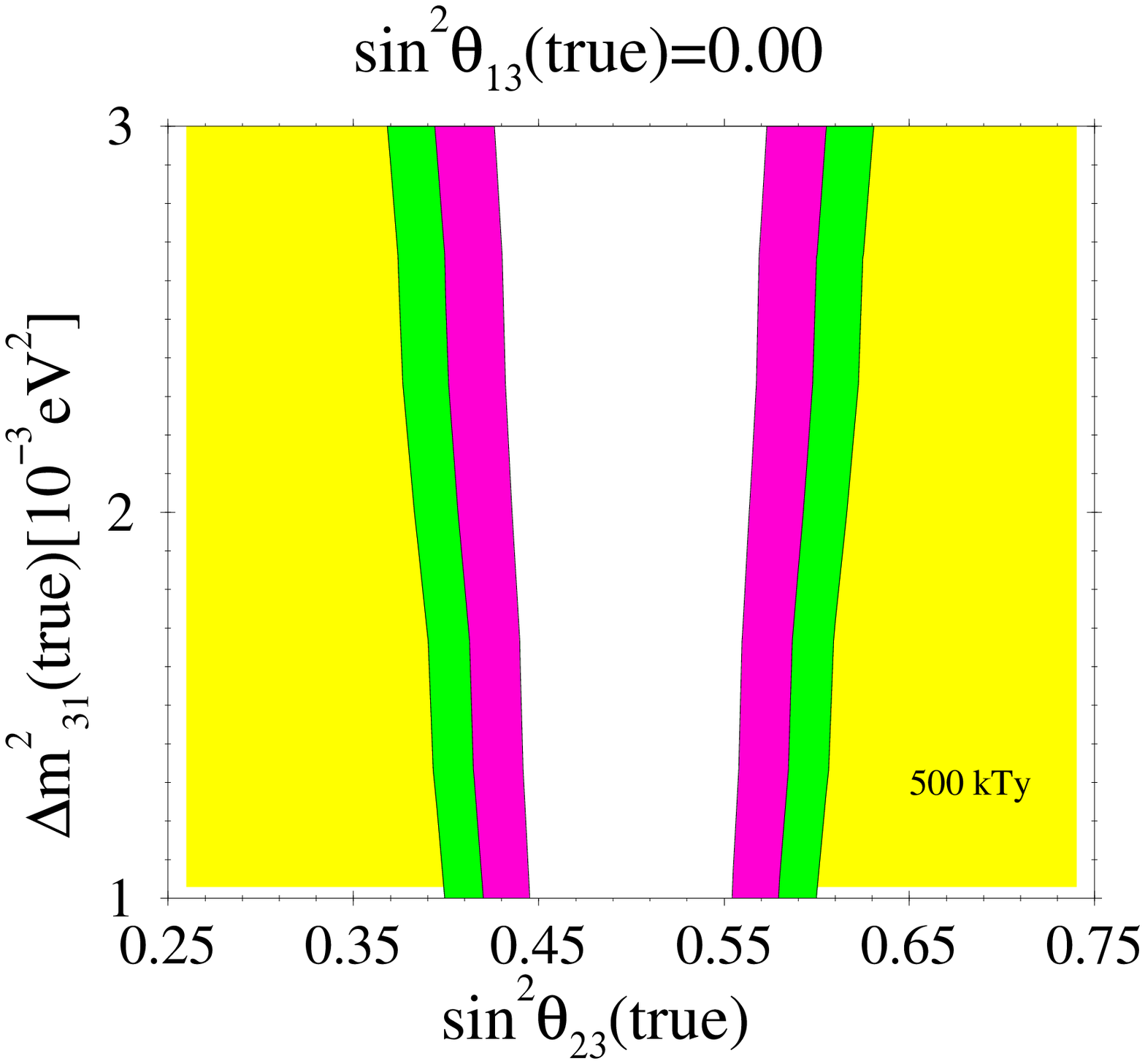}
\caption{\label{fig:maximality}
The regions of $\Delta m_{31}^2$(true) and 
$\sat$ where maximal $\theta_{23}$
mixing can be rejected 
at $1\sigma$ (inner bands), $2\sigma$ (middle bands)
and $3\sigma$ (outer bands) C.L. 
The left-hand panel \cite{Antusch:2004yx}
shows the 
sensitivity expected from the combined data from the LBL
experiments. The middle panel \cite{Gonzalez-Garcia:2004cu}
shows the sensitivity expected with atmospheric neutrinos 
in a megaton water 
detector (SK50). The extreme right-hand panel \cite{Choubey:2005zy}
shows the corresponding reach expected from 500 kTy 
atmospheric neutrino data in large magnetized iron detectors.
The true value of $\theta_{13}$ is assumed to be zero.
}
\end{figure}
%%%%%%%%%%%%%%%%%%%%%%%%%%%%%%%%%%%%%%%%%%%%%%%
%%%%%%%%%%%%%%%%%%%%%%%%%%%%%%%%%%%%%%%%%%%%%%%%%%%%%%%%%%
\begin{figure}
\includegraphics[width= 5.0cm, height=5.2cm]{huber_maxmix_1panel.ps}%\\
\vglue -5.2cm \hglue 5.0cm
\includegraphics[width= 5.0cm, height=5.0cm]{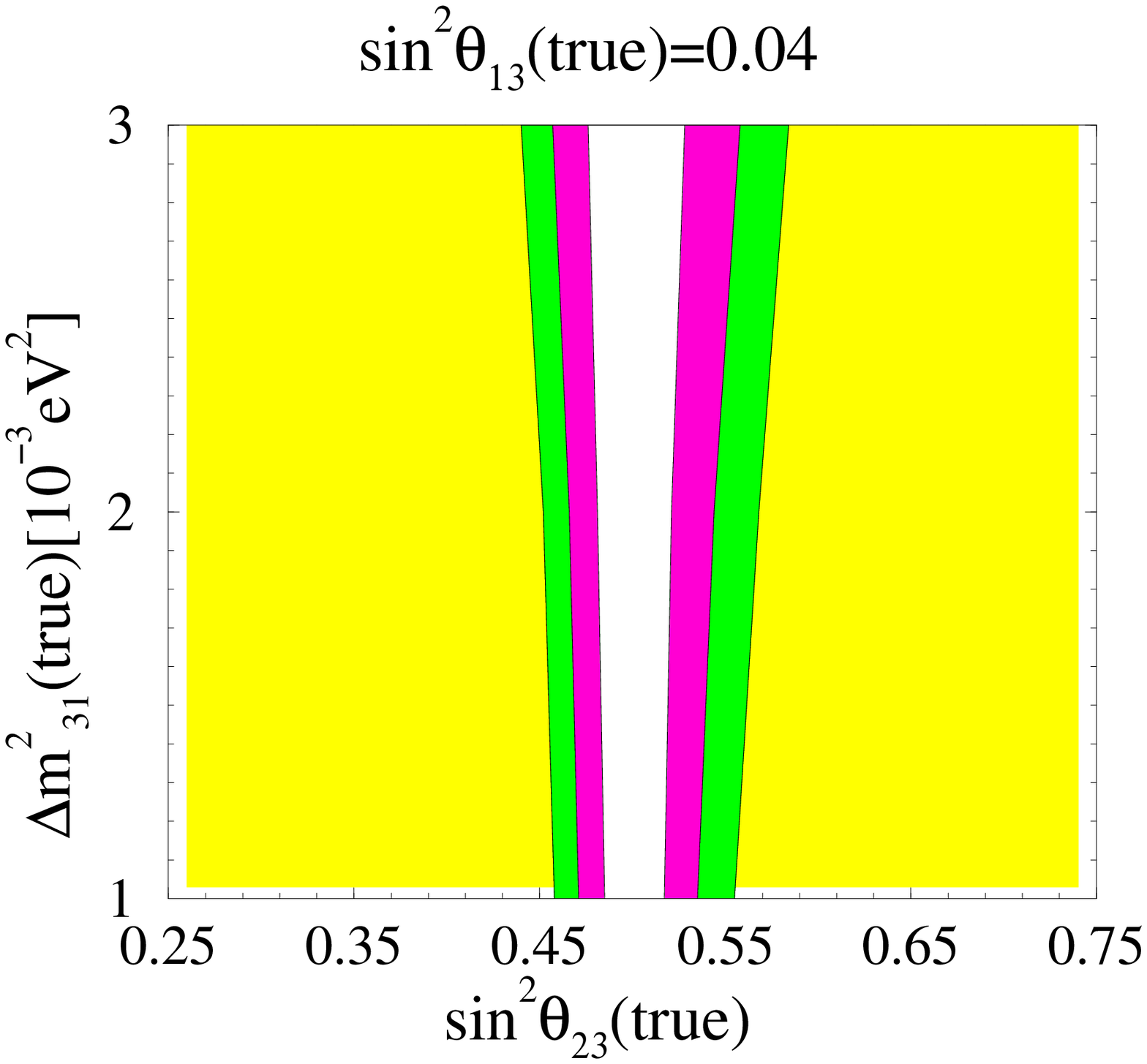}\\
\vglue -5.5cm \hglue 10.5cm
\includegraphics[width= 5.0cm, height=5.0cm]{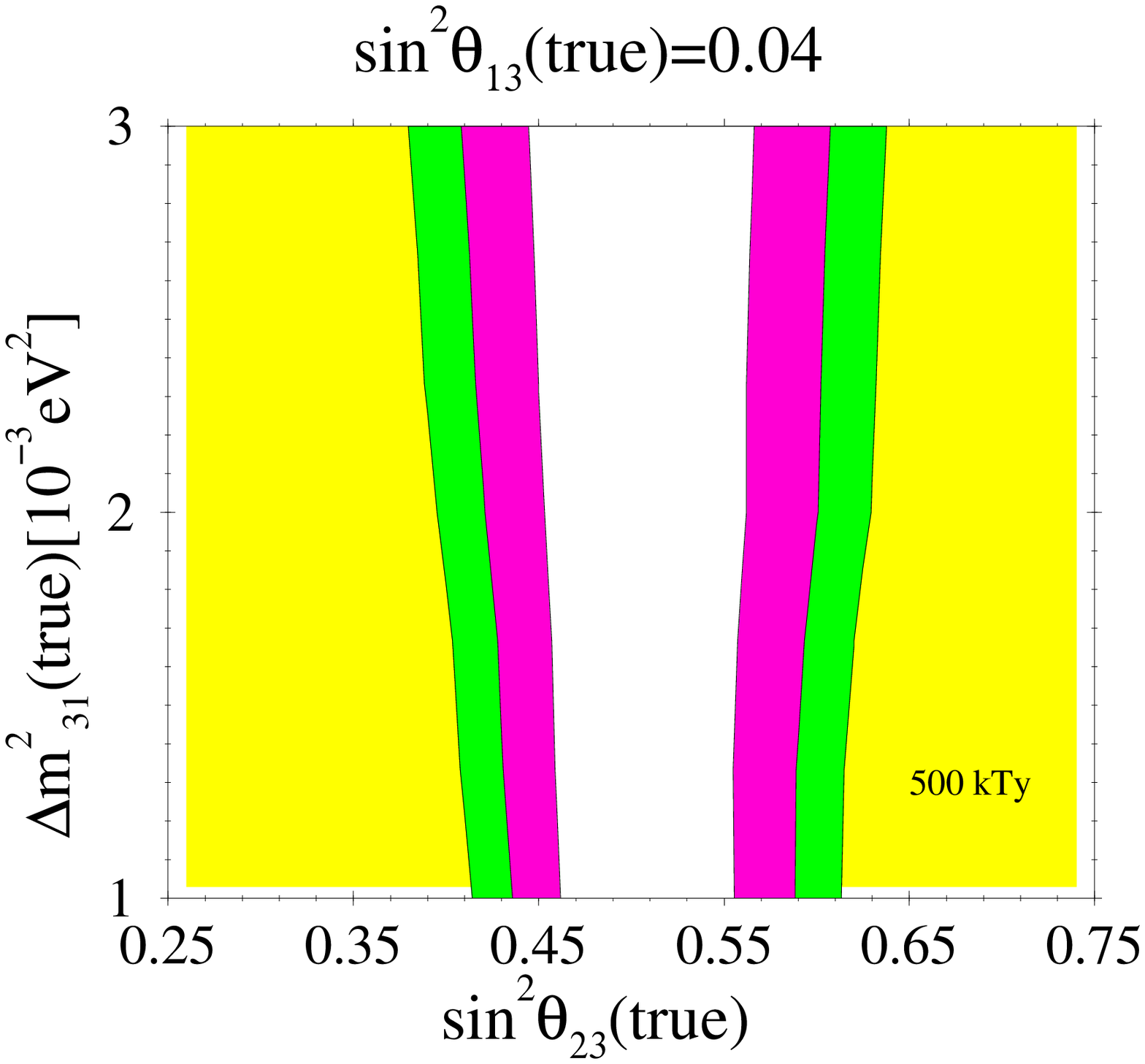}
\caption{\label{fig:maximality13}
Left-hand panel same as in Fig. \ref{fig:maximality}. Middle panel 
(for water detector)
and right-hand panel (for magnetized iron detector) 
have been drawn assuming that  
$\scht=0.04$.}
\end{figure}
%%%%%%%%%%%%%%%%%%%%%%%%%%%%%%%%%%%%%%%%%%%%%%%

The measurement of both the magnitude and sign of   
the deviation of $\sa$ from its maximal value 0.5 is of 
utmost theoretical importance. 
To quantify the deviation of the true value of $\theta_{23}$ from 
its maximal value, we introduce the function
$D \equiv \frac{1}{2} - \sa.$
The magnitude $|D|$ gives the deviation of $\sa$ from its maximal 
value, while $sgn(D)$ gives the octant of $\sa$.
The best current limit on 
$|D|$ comes from the SK atmospheric neutrino 
experiment giving $|D| \leq 0.16$ at $3\sigma$ \cite{Ashie:2005ik}
while $sgn(D)$ is almost unknown at present.
Fig. \ref{fig:maximality} shows the potential of atmospheric 
neutrino experiments to test the deviation of $\theta_{23}$ from 
maximality and compares it with the reach of the combined 
data from the current and next generation long baseline 
experiments. The combined long baseline data 
include five years of running each of the MINOS, ICARUS, OPERA, 
T2K and NO$\nu$A experiments. The middle panel gives the 
sensitivity to $|D|$ of  
atmospheric neutrino experiments with 
water detectors for a 4.6 Megaton-yr statistics, while
the left panel shows the corresponding sensitivity of 
atmospheric neutrino data in large magnetized iron detectors
with 500 kTy statistics. 
At $\Delta m_{31}^2$(true)$=2.5\times 10^{-3}$ eV$^2$, it should be 
possible to measure $|D|$ within 19\% and 25\% at $3\sigma$
with atmospheric neutrinos in water and iron detectors respectively. 
This is slightly weaker than the sensitivity 
of the combined 
long baseline experiments, where it should be possible 
to measure $|D|$ 
within 14\% at $3\sigma$. 
However, note that all the results presented in 
Fig. \ref{fig:maximality} have been obtained assuming that the true 
value of $\theta_{13}$ was zero. Results for 
atmospheric neutrino experiments when the true value of  
$\theta_{13}$ is not zero is shown in Fig. \ref{fig:maximality13}.
For non-zero $\theta_{13}$, 
presence of earth matter effects in the $P_{\mu\mu}$ channel 
brings in a marginal improvement in the sensitivity of atmospheric 
neutrino experiment with the magnetized iron detector. For the 
megaton water atmospheric neutrino experiment, 
very large earth matter effects in the $P_{\mu e}$ channel  
bring in significant improvement in the determination 
of $|D|$, making this experiment comparable/better than the 
long baseline experiments for studying the deviation of 
$\theta_{23}$ from maximality.

%%%%%%%%%%%%%%%%%%%%%%%%%%%%%%%%%%%%%%%%%%%%%%%%%%%%%%%%%%%%%%%%
\section{\label{sec:octant}
Resolving the $\theta_{23}$ Octant Ambiguity} 
%%%%%%%%%%%%%%%%%%%%%%%%%%%%%%%%%%%%%%%%%%%%%%%%

%%%%%%%%%%%%%%%%%%%%%%%%%%%%%%%%%%%%%%%%%
\begin{figure}
%\begin{center}
\hglue 1cm
\includegraphics[width=6.0cm, height=5.5cm]
{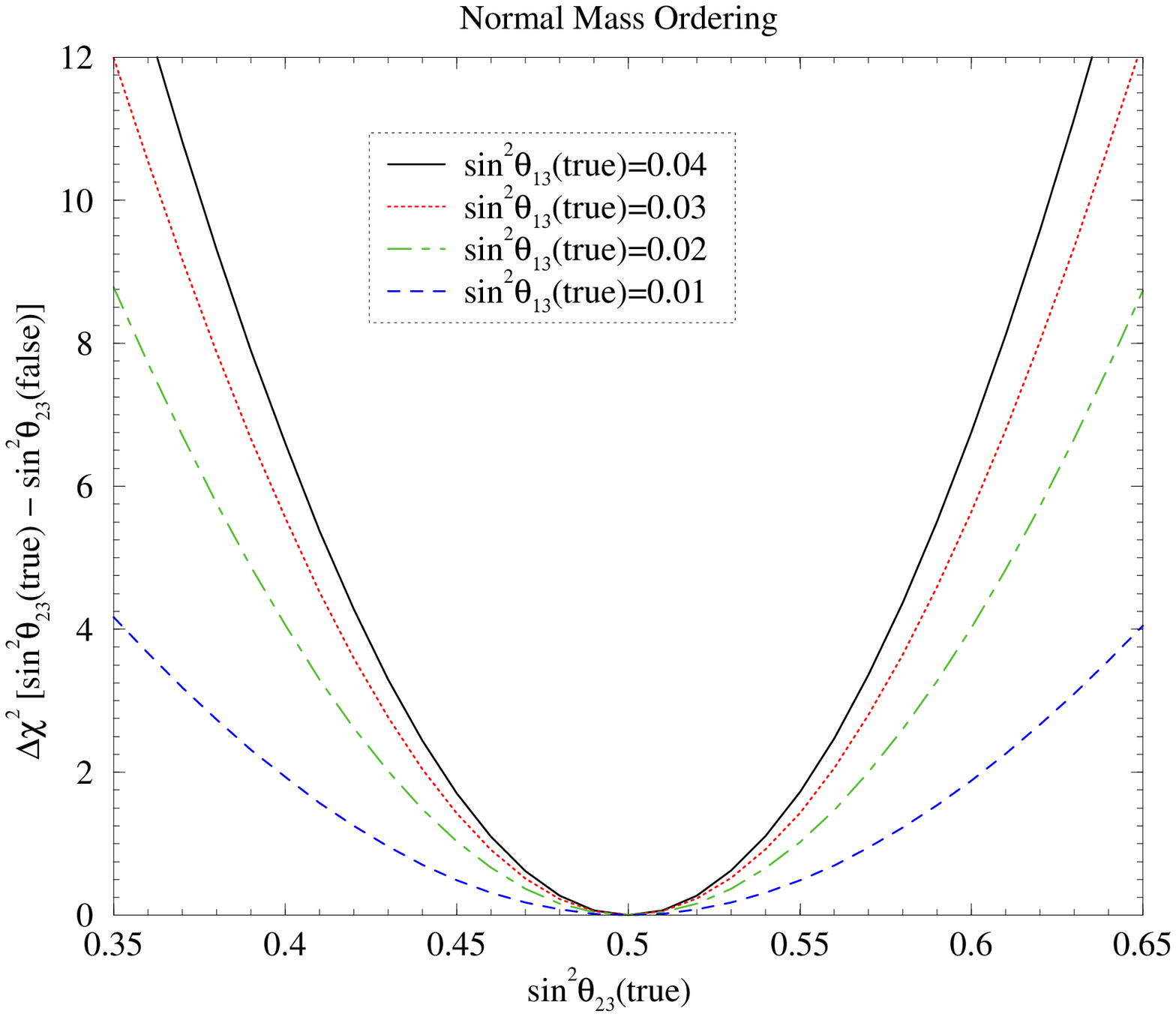}
\vglue -5.5cm \hglue 7.8cm
\includegraphics[width=6.0cm, height=5.5cm]
{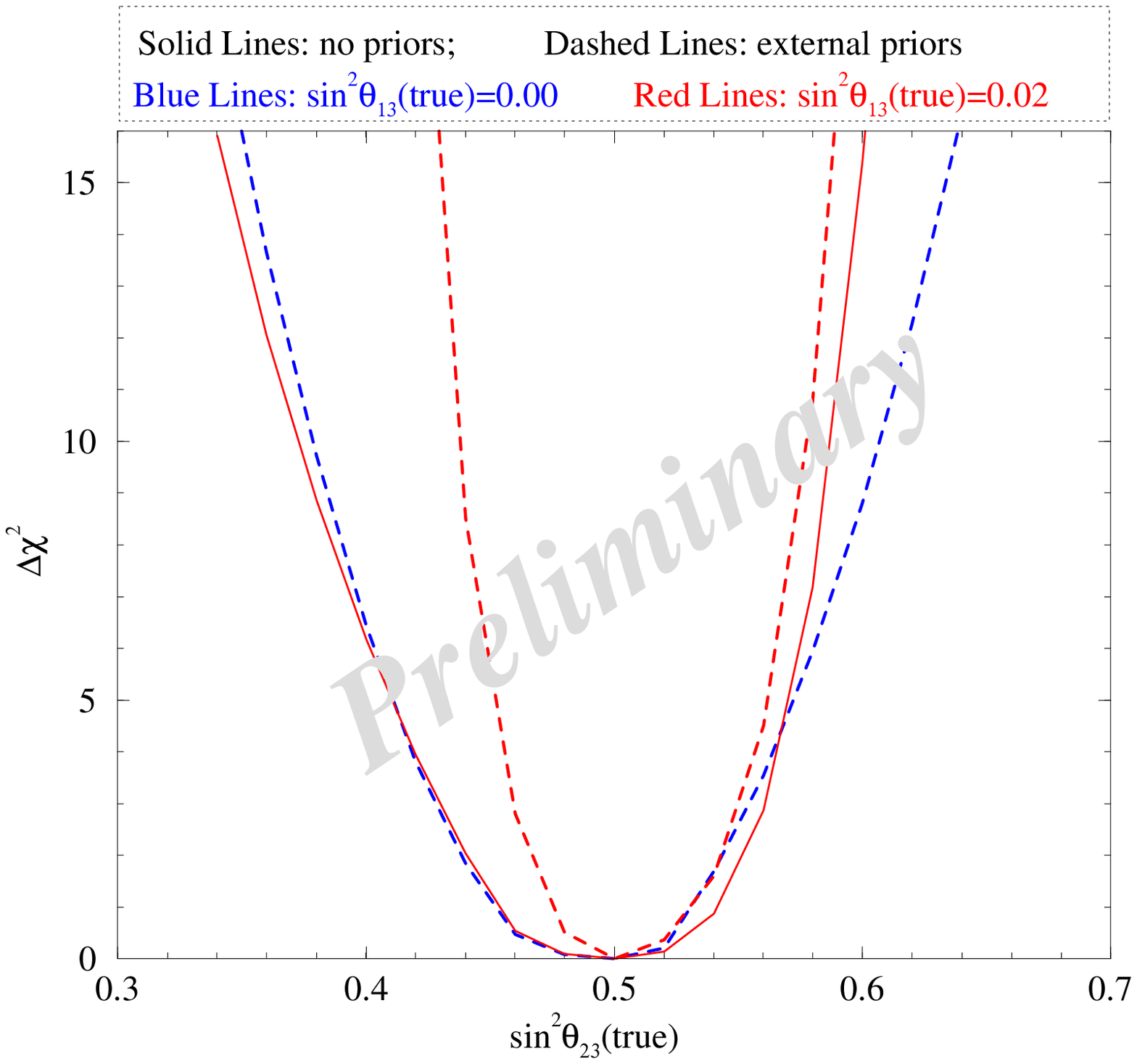}
\caption{Plot showing the octant sensitivity 
as a function of $\sin^2\theta_{23}$(true), for 
an atmospheric neutrino experiment with 
large magnetized iron calorimeter (left-hand panel)
and megaton water detector (right-hand panel).
}
\label{fig:delchioctant}
%\end{center}
\end{figure}

\begin{table}
\begin{tabular}{ccc}
\hline
%& &  \cr
Type of Experiment & $\sin^2\theta_{23}$(false) excluded
at 3$\sigma$ if: & for \cr
%& &  \cr
\hline
%& &  \cr
%&$\sin^2\theta_{23}({\rm true}) < 0.361 \ {\rm or} \ > 0.633$
%& $\sin^2\theta_{13} ({\rm true}) = 0.01$ \cr
&$\sin^2\theta_{23} ({\rm true}) < 0.402 \ {\rm or} \ > 0.592 $
&$\sin^2\theta_{13} ({\rm true}) = 0.02$ \cr
Magnetized Iron (0.5 MTy)
%&$\sin^2\theta_{23} ({\rm true}) < 0.415 \ {\rm or} \ > 0.580 $
%&$\sin^2\theta_{13} ({\rm true}) = 0.03$ \cr
&$\sin^2\theta_{23} ({\rm true}) < 0.421 \ {\rm or} \ > 0.573$
&$\sin^2\theta_{13} ({\rm true}) = 0.04$ \cr
%& &  \cr
\hline
%& &  \cr
& 
$\sin^2\theta_{23} ({\rm true}) < 0.383 \ {\rm or} \ > 0.600$
& $\scht=0.00$ \cr
Water Cerenkov (4.6 MTy) & 
$\sin^2\theta_{23} ({\rm true}) < 0.438 \ {\rm or} \ > 0.573$
& $\scht=0.02$ \cr
%& &  \cr
\hline
\end{tabular} 
\caption{\label{tab:octant}
A comparison of the potential of different experiments 
to rule out the wrong $\theta_{23}$ octant at \sig{} (1 dof). The third
column gives the condition on the true value of $\sch$ needed 
for the $\theta_{23}$ octant resolution.
}
\end{table}

If the true value of $\theta_{23}$ is not $45^\circ$, then 
the question arises whether $\theta_{23} >$ ($D$ positive)  
or $< \pi/4$ ($D$ negative). This leads to an 
additional two-fold degeneracy in the 
measurement of the mixing angle $\theta_{13}$ and the CP phase 
$\delta_{CP}$ in LBL experiments. This ambiguity is generally 
regarded as the most difficult to resolve.
As discussed before, the presence of earth matter 
effects in the zenith angle and energy binned 
atmospheric $\numu/\anumu$ data in magnetized iron detectors 
opens up the possibility of probing 
the octant of $\theta_{23}$ \cite{Choubey:2005zy}.
On the other hand atmospheric $\nue/\anue$ data
in water detectors could also give information on the octant 
of $\theta_{23}$, both through the $\Delta m_{21}^2$ dependent 
subdominant term in the sub-GeV sample 
\cite{Peres:1999yi,Gonzalez-Garcia:2004cu},
as well as through earth matter effect in the multi-GeV events,
%\cite{SC:nu2006talk,Huber:2005ep}, 
as discussed above. One could hence combine
%This therefore opens up the 
%possibility of combining 
the atmospheric neutrino data,  
in either megaton water detectors \cite{Huber:2005ep,Campagne:2006yx}, 
or in large magnetized iron calorimeters
with data from long baseline experiments to resolve 
parameter degeneracies.

In order to obtain the limiting value of $\sin^2\theta_{23}$(true) 
which could still allow for the determination of $sgn(D)$
we define
\be
\Delta \chi^2 \equiv \chi^2 (\sin^2\theta_{23} ({\rm true}),
\sin^2\theta_{13} ({\rm true}), {\rm others(true)}) 
- \chi^2(\sin^2\theta_{23} ({\rm
false}),\sin^2\theta_{13}, {\rm others}),
\label{Eq:chioctant}
\ee
with $\sin^2\theta_{23}$(false) restricted to the wrong octant and 
`others' comprising $\Delta m^2_{31}$, $\Delta m^2_{21}$,
$\sin^2\theta_{12}$ and $\delta_{CP}$.  These, along with
$\sin^2\theta_{13}$ as well as $\sin^2\theta_{23}$(false), 
are allowed to vary freely in the fit.  
Fig. \ref{fig:delchioctant} 
shows the results 
of a statistical analysis based on simulated data from 
atmospheric neutrinos with 500 kiloton-yr
exposure in a large magnetized iron calorimeter
(left-hand panel) and 4.6 Megaton-yr exposure in 
a water Cerenkov experiment (right-hand panel). For 
large magnetized iron detector we show results for four 
different values of $\sin^2\theta_{13}$(true), assuming a 
normal mass ordering. 
For a given $\sin^2\theta_{13}$(true), the range of
$\sin^2\theta_{23}$(true), for which $\sin^2 \theta_{23}$(false) can
be ruled out with atmospheric neutrinos in magnetized iron 
detector is given in Table \ref{tab:octant}. This can be 
compared to the sensitivity possible with water Cerenkov 
detectors, shown for a true 
normal hierarchy in the right-hand panel of 
Fig. \ref{fig:delchioctant} and Table \ref{tab:octant}, 
where octant determination can be done reasonably well
even if $\sin^2\theta_{13}$(true) was zero \cite{Gonzalez-Garcia:2004cu}.
However, 
if $\sin^2\theta_{13}$(true) was non-vanishing and reasonably
large, the octant sensitivity of this experiments gets 
significantly boosted through earth matter effects appearing 
in the multi-GeV electron sample.

%%%%%%%%%%%%%%%%%%%%%%%%%%%%%%%%%%%%%%%%%%%%%%%%%%%%%%%%
\section{Resolving the Ambiguity in Neutrino Mass Hierarchy}
%%%%%%%%%%%%%%%%%%%%%%%%%%%%%%%%%%%%%%%%%%%

\begin{figure}
\begin{center}
\includegraphics[width=16.0cm, height=5.5cm]
{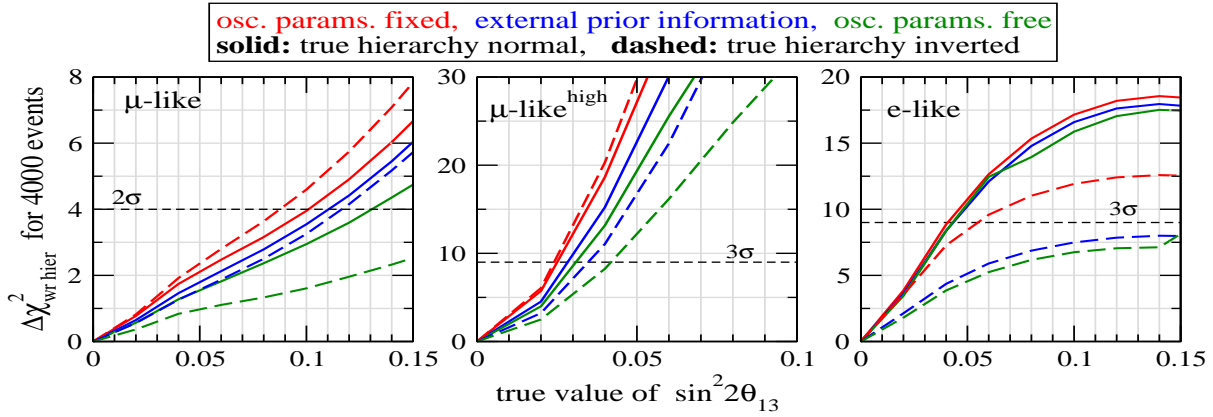}
\caption{
$\Delta \chi^2$ for the wrong hierarchy as a function of 
$\sin^22\theta_{13}$(true). See \cite{Petcov:2005rv} for details.
}
\label{fig:hierarchyINO}
\end{center}
\end{figure}

Large earth matter effects in atmospheric neutrinos can be 
exploited to probe the sign of $\Delta m_{31}^2$. 
Fig. \ref{fig:hierarchyINO} \cite{Petcov:2005rv}
shows the sensitivity to
$sgn(\Delta m_{31}^2)$ expected in a magnetized iron 
calorimeter, with 4000 observed upward going events.
The data corresponds to a normal (solid lines) and 
inverted (dashed lines) hierarchy and the curves show the 
$\chi^2$ and hence the C.L. 
with which the wrong hierarchy can be ruled out. 
The red lines correspond to an analysis method where all the 
oscillation parameters are fixed in the fit. The blue lines 
show the results of the fit where external priors for 
the oscillation parameters have been used. The green lines 
correspond to the case where all oscillation parameters are allowed 
to vary freely in the fit. 
The left-hand panel is for muon events in 
a detector with 
15\% energy and $15^\circ$ zenith angle resolution, the 
middle panel is for muon events with 5\% energy and $5^\circ$ 
zenith angle resolution, while the 
right-hand panel is for electron events.
For vanishing $\theta_{13}$ the matter effects vanish 
giving $\chi^2=0$. As $\theta_{13}$ increases, matter effects 
increase, thereby increasing the sensitivity of the 
experiment to hierarchy determination. For a INO-ICAL like 
detector, where energy resolution is expected to be around 15\%
and zenith angle resolution of about $15^\circ$, 
the wrong hierarchy can be ruled out at 
$2\sigma$ using the muon events, 
if
$\sin^22\theta_{13}$(true)$=0.1$ and $\sin^2\theta_{23}$(true)$=0.5$,
and where the information from the other long baseline experiments 
on the oscillation parameters have been included through the priors. 
Comparison of the left-hand with the middle panel 
shows that the sensitivity to hierarchy increases if the 
detector resolution is improved. 
Comparison of the left-hand with the right-hand panel 
shows that the sensitivity to hierarchy increases if the 
detector could detect electron type events as well. And of course
since matter effects increase with $\theta_{23}$, the 
sensitivity to hierarchy increases as the true value of
$\theta_{23}$ increases.

The $sgn(\Delta m_{31}^2)$ can be done using the excess in the 
multi-GeV electron sample due to earth matter effects in water 
Cerenkov detectors.
The wrong hierarchy can be ruled by a 4.6 Megaton-yr data 
in such an experiment at more than the 2$\sigma$ limit if 
$\sin^22\theta_{13}$(true)$=0.1$ and $\sin^2\theta_{23}$(true)$=0.5$
(see also \cite{Huber:2005ep}).
This is comparable to the sensitivity of the magnetized iron 
detectors as discussed above. However, since water detectors use 
the excess in electron events for multi-GeV neutrinos, which 
in turn have large matter effects in the $P_{\mu e}$ channel,
they therefore depend also on the CP phase $\delta_{CP}$ as 
discussed before. If the value of $\delta_{CP}$ is allowed to vary 
freely in the fit then the sensitivity gets affected and 
decreases appreciably.

%%%%%%%%%%%%%%%%%%%%%%%%%%%%%%%%%%%%%%%%%%%%%%%%%%%%
%%%%%%%%%%%%%%%%%%%%%%%%%%%%

\section{Looking for new physics}
There are a variety of new physics scenarios which could manifest
themselves as subdominant effects in oscillations of atmospheric 
neutrinos. Each one of these have a distinctive $L/E$ behavior, 
while oscillations go linearly with $L/E$. 
%The very wide range of 
%$L/E$ can therefore be used very effectively to probe/constrain 
%the new physics parameters. 
This can be probed directly using the 
$L/E$ binned data in large iron detectors \cite{cpt}
or by comparing the 
low energy contained events with the high energy upward 
going muons in large water detectors \cite{sknewphy}. 
Atmospheric neutrino 
events which constitute a background for the neutrino telescopes
such as IceCube, can also be used every effectively to constrain 
new physics. 
Neutrino telescopes
look for neutrinos in the
$(10^{-1}-10^4)$ TeV range, for which standard 
oscillations of atmospheric neutrinos 
are negligible. Hence, any $E$ or $L$ dependence in 
the data would signify new physics. 
Likewise, the absence of any 
$E$ and $L$ dependence can be used to constrain the new physics 
parameters \cite{newatm}.

\section{Conclusions}
In this talk we explored the 
physics potential of future measurements
of atmospheric neutrinos. Observation of 
$\Delta m^2_{21}$ driven sub-dominant 
effects and $\theta_{13}$ driven 
large matter effects in atmospheric neutrinos can be 
used to study the 
deviation of $\theta_{23}$ from maximality and its octant.
Neutrino mass hierarchy can be determined extremely well due 
to the large matter effects. New physics can be constrained
both in standard atmospheric neutrino experiments as well as 
in future neutrino telescopes.

\vglue 0.5cm
%%%%%%%%%%%%%%%%%%%%%%%%%%%%%%%%%%%%%%%%%%%%%%%%%%%%%%%%%%%%%%%

\end{document}